\documentclass[review]{elsarticle}

\usepackage{multirow}
\usepackage{booktabs}  
\usepackage{makecell}
\usepackage{graphicx}
\usepackage{subfigure}
\usepackage{float}

\begin{document}

\begin{frontmatter}

\title{A SPA-based Manifold Learning Framework for Motor Imagery EEG Data Classification\tnoteref{mytitlenote}}

\author{Xiangyun Li\textsuperscript{a}}
\author{Peng Chen\textsuperscript{b}\corref{mycorrespondingauthor}}



\cortext[mycorrespondingauthorr]{Corresponding author}
\ead{chenpeng@swjtu.edu.cn}

\author{Zhanpeng Bao\textsuperscript{b}}

\address[mymainaddress]{West China Biomedical Big Data Center, West China Hospital, Sichuan University,Chengdu 610041, PR China}
\address[mysecondaryaddress]{School of Mechanical Engineering, Southwest Jiaotong University, Chengdu 610031, PR China}

\begin{abstract}
The electroencephalography (EEG) signal is a non-stationary, stochastic, and highly non-linear bioelectric signal for which achieving high classification accuracy is challenging, especially when the number of subjects is limited. As frequently used solution, classifiers based on multilayer neural networks has to be implemented without large training data sets and careful tuning. This paper proposes a manifold learning framework to classify two types of EEG data from motor imagery (MI) tasks by discovering lower dimensional geometric structures. For feature extraction, it is implemented by Common Spatial Pattern (CSP) from the preprocessed EEG signals. In the neighborhoods of the features for classification, the local approximation to the support of the data is obtained, and then the features are assigned to the classes with the closest support. A spherical approximation (SPA) classifier is created using spherelets for local approximation, and the extracted features are classified with this manifold-based method. The SPA classifier achieves high accuracy in the 2008 BCI competition data, and the analysis shows that this method can significantly improve the decoding accuracy of MI tasks and exhibit strong robustness for small sample datasets. It would be simple and efficient to tune the two-parameters classifier for the online brain-computer interface(BCI)system.
\end{abstract}

\begin{keyword}
\texttt Motor imagery electroencephalography \sep Brain-computer interface \sep Manifold learning \sep Local manifold approximation \sep Classification algorithm
\end{keyword}

\end{frontmatter}


\section{Introduction}
Electroencephalography (EEG)-based brain-computer interface (BCI) technology is a promising tool that enables users to use their brain activity to directly control or interact with an external environment or device \cite{chaudhary2016brain}, especially for persons with severe motor-related diseases \cite{lazarou2018eeg}. Despite its low spatial resolution, EEG is the most widely used technique in the field of BCI due to its good temporal resolution \cite{khan2014decoding}, low equipment cost, and high compatibility \cite{wang2006common}. The BCI technique has been studied via EEG signals arising in different physiological mechanisms, such as motor imagery (MI) \cite{brunner2007spatial}, steady-state visual evoked potentials (SSVEP) \cite{lin2006frequency}, and P300 \cite{krusienski2008toward}. Compared with BCI based on SSVEP or P300, MI methods may have higher potential because they are independent of external stimuli, thus allowing asynchronous control and communication, and facilitating the application of BCI system \cite{naseer2015decoding}. Although there are slight differences of EEG patterns between, recent advances in this field show that patients can produce a cortical activation similar to healthy individuals during the imagined movements \cite{foldes2017altered}. Therefore, the MI-based BCI system is an effective way to promote recovery during rehabilitation interventions, which is particularly suitable for tetraplegic individuals to restore neuromuscular connectivity and motor function \cite{pichiorri2015brain}. The goal of MI-based system is to provide a neurophysical communication channel between people and external devices.For safety purpose of related technologies, noninvasive EEG-based BCIs are widely used, in such applications as word spellers \cite{martens2010generative}, wheelchair control \cite{galan2008brain}, and video games \cite{tangermann2008playing}. In addition, non-invasive BCI may also be used in evaluating the brain activity of severely paralyzed patients to predict the efficiency of invasive brain-computer interfaces \cite{fukuma2015closed}. For healthy persons, BCI can also greatly enhance the experience of multimedia and video games and thus have have vast commercial value \cite{nijholt2009turning}.

The major challenges in a typical MI task are the effective extraction of EEG features and and their proper classification. EEG signals is usually represented as high dimensional data whose classification is prone to overfitting and bias, especially for small training datasets \cite{jackson2001adaptive}. Dimensionality reduction is an effective method to solve these problems for obtaining a more compact representation of the high-dimensional space. Manifold learning \cite{freedman2002efficient} is one of the most important nonlinear dimensionality reduction techniques, 
and the goal of most manifold learning algorithms is to obtain low-dimensional embedding of high-dimensional data so that the proximal data points from the high-dimensional space remain close to each other as in the high-dimensional space, so are the distant data points \cite{xie2016motor}.

With the advantages promoting the classification performance with the limited data settings due to its simple and direct mathematical representations, many studies have recently proposed manifold learning algorithms for BCI applications,and EEG classification on high-dimensional Riemannian manifold has recently received increasing attention \cite{barachant2010riemannian}. Barachant et al. \cite{barachant2011multiclass} proposed two classification algorithms, one of which introduces the concept of Riemannian geodesic distance to compare the minimum Riemannian distance between unlabeled data points and the Riemannian mean of labeled data points. The other algorithm maps all data points in the Riemannian manifold into its tangent space (TS),called the optimal hyperplane for classification \cite{tuzel2008pedestrian},and then applies the linear classification method on the tangent space. It is different from the method of global sampling manifold information in reference \cite{barachant2011multiclass}. Xie et al used a simple but effective bilinear sub-meridian learning (BSML) algorithm \cite{xie2016motor} in which the intrinsic sub-meridians are learned by identifying a bilinear mapping that maximally preserves the local geometry and overall structure of the original manifold for dimensionality reduction of the SPD matrix space in the motor imagery BCIs. Li et al. \cite{li2019classification} proposed a local manifold approximation classification, which is mainly based on obtaining a local approximation to the support of the data in each class within the neighborhood of the feature to be classified, and assigning the feature to the class with the closest support. The classifier performs a local approximation using easily fitted spheres to result in a spherical approximation (SPA) classifier.

Unlike general machine learning tasks, classification tasks for BCI often lack sufficient labeled training data because of the limited number of subjects. In practice, It is usually that case there is an insufficient number or even no training samples at all, known as the small training data set problem. If the amount of available data is small, the model is difficult to generalize, and the classification accuracy tends to be lowered \cite{jackson2001adaptive}. There is significant variation in EEG signals between different individuals, characterized by the inconsistent distribution of data in feature space, which prevents prevent the model transferred from one person to other \cite{craik2019classification}. Therefore, designing an algorithm with high classification accuracy constrained by small training data sets is a challenging task in the current BCI field. 

The innovations and contributions of this paper include the following three aspects:
\begin{enumerate}[(1)]
\item The SPA classifier based on manifold learning is modified and applied to the BCI system for decoding the features of MI-based EEG signals. In the case of no geodesic information of the unknown manifold, SPA uses sphere for local approximation, which is different from other manifold learning methods that approximate geodesic distance from the global point of view. The sphere is a simple geometric object that is easy to fit, and it also provides the improvements of the hyperplane, so that the accuracy can be promoted by approximating the local curvature. It is the first application of SPA in the classification of EEG signals, especially for MI data.
\item Compared with the usual algorithms for a large number of training samples, This algorithm is suitable for the system such as BCI which lacks enough training data, and can still present effective feature decoding performance of EEG signals for a small number of samples with the SPA classifier, as demonstrated by the experimental results. In clinical rehabilitation training, such a classifier could work with the limited number of subjects.
\item The design of the SPA classifier is simple, and there are only two parameters needed to be tuned. The main parameters need to be adjusted are the size of the local neighborhoods K and the dimension of the manifold approximating P. After these two parameters are determined, the algorithm execution will become very fast, and become suitable for online BCI systems requiring high real-time performance.
\end{enumerate}

The rest of this article is divided into the following sections. Section 2 describes the MI task experimental example and basic information on the EEG data set for the 2008 BCI competition, as well as the preprocessing methods for EEG signals, and presents the Common Space Pattern Feature Extraction algorithm for MI-based classification problems. Section 3 describes the SPA classifier based on manifold learning in detail. In Section 4, the experimental results of this algorithm and other commonly used algorithms are compared for evaluation and discussion.In the end, Section 5 provides conclusions and outlines the future applications of SPA classifiers.

\section{Motor imagery EEG signals classification preparation}

\subsection{Introduction to the BCI competition dataset}

The dataset used in this paper is from the BCI Competition IV Dataset 2a, which consisted of four MI tasks: motor imagery of the left hand, right hand, both feet, and tongue for nine subjects. EEG signals from 22 Ag/AgCl electrodes and 3 monopolar electrooculography (EOG) channels (with the left mastoid as reference) were recorded at a sampling frequency of 250 Hz. The EEG signal is band-pass filtered between 0.5 and 100 Hz, and the power line interference is filtered by a 50 Hz notch filter. The timing scheme of the experimental paradigm is shown in Figure 1. More detailed information about the EEG experiment can be found in \cite{tangermann2012review}. Taking into account the non-stationary of the EEG data, the EEG data consists of two sessions, recorded on different dates. Each session contains a total of 288 trials, and each type of task is executed 72 times. It indicates that the dataset includes contaminated data and such data contains some artifacts. However, these contaminated samples are usually not removed to examine the robustness of the proposed algorithm. In the experiment, only the data related to the motor imagery of the left hand and right hand were extracted from the dataset, while the motor imagery data of the tongue and feet were excluded. 

\begin{figure}
\centering
\includegraphics[width=4.00in,height=1.50in]{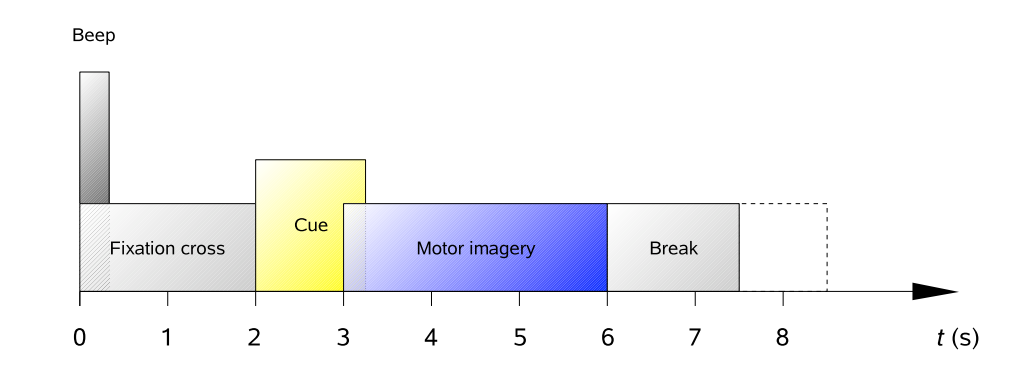}
\caption{BCI competition IV dataset 2a experimental paradigm}
\label{1} 
\end{figure}

\subsection{EEG data preprocessing process}

\subsubsection{Time Interception}

The MI task requires subjects to repeatedly imagine limb movements for a certain amount of time to generate stable and effective brain activity \cite{asensio2013extracting}. For the continuously recorded EEG signals, it only focuses on the motor imagery part and then segments the band-pass filtered continuous EEG signals through the time window. The segmented time length corresponds to a single motor imagery trial. Besides, this study used signals between 500 ms and 2500 ms after the stimulus occurred and processed only the intercepted segments.

\subsubsection{Band-pass filter}

During motor imagery, event-related synchronization and desynchronization (ERS/ERD) phenomena occur mainly in the 8-30 Hz band, covering both mu (8-13 Hz) and beta (14-28 Hz) rhythms \cite{pfurtscheller2006mu}. Typically, a bandpass filter is used to highlight the frequency components of interest while reducing some artifacts. In this study, a fifth-order Butterworth bandpass filter was designed, which eliminates most of the muscle activity, such as EMG (above 35 Hz) caused by swallowing or facial movements. It also removes some ocular artifacts that mainly affect the low-frequency components of the EEG signal, such as the EOG (below 8 Hz) generated by blinking and motion.

\subsection{Common Spatial Pattern (CSP) for Feature Extraction}

At present, the spatial domain-based feature extraction method has become the main feature extraction method for MI EEG signals, which is represented by the common spatial pattern (CSP). CSP was first proposed by Fukunaga in 1970 and then introduced into the analysis and processing of EEG signals by Koles \cite{koles1990spatial}, and it has made a very large number of applications on MI EEG signals \cite{ramoser2000optimal}. CSP is a supervised learning method, which requires the training data to be labeled. CSP is mainly a feature extraction algorithm for two classification tasks, capable of extracting the spatial distribution components of each class from multi-channel brain-computer interface data.

Suppose $X_1$ and $X_2$ are the multi-channel evoked response time-space signal matrices under the binary classification MI tasks, respectively. The dimensions of $X_1$ and $X_2$ are all N*T, where N is the number of EEG channels and T is the number of samples collected per channel.

To calculate its covariance matrix, assume that N\textless T. The normalized covariance matrices $R_1$ and $R_2$ of $X_1$ and $X_2$ are:

\begin{equation}
R_1=(X_1 X_1^T)/(trace(X_1 X_1^T)) R_2=(X_2 X_2^T)/(trace(X_2 X_2^T))
\end{equation}

Then, the mixed space covariance matrix R is calculated by:
\begin{equation}
    R=\overline{R_{1}}+\overline{R_{2}}
\end{equation}

Perform the eigenvalue decomposition on the mixed space covariance matrix R:
\begin{equation}
    R=U \lambda U^{T}
\end{equation}

Arrange the eigenvalues in descending order, and calculate the whitening value matrix P:
\begin{equation}
    P=\sqrt{\lambda^{-1}} U^{T}
\end{equation}

Then, the normalized covariance matrices $R_1$ and $R_2$ are transformed as follows:
\begin{equation}
    S_{1}=P R_{1} P^{T} \quad S_{2}=P R_{2} P^{T}
\end{equation}

Next, do the principal component decomposition for $S_1$ and $S_2$:
\begin{equation}
    S_{1}=B_{1} \lambda_{1} B_{1}^{T} \quad S_{2}=B_{2} \lambda_{2} B_{2}^{T}
\end{equation}

The above formula proves that the eigenvector matrices of matrices $S_1$ and $S_2$ are equal, i.e., $B_1$=$B_2$. Sort the eigenvalues in $\lambda_{1}$ in descending order, and the corresponding eigenvalues in $\lambda_{2}$ in ascending order. The transformation of the whitened EEG signal with the eigenvectors corresponding to the largest eigenvalues in $\lambda_{1}$ and $\lambda_{2}$ is optimal for separating the variances in two signal matrices. The spatial filter corresponding to the projection matrix W is:
\begin{equation}
    W=B^{T} P
\end{equation}

For the test data $X_i$, the feature vector f is extracted as follows:
\begin{equation}
    \left\{\begin{array}{c}
        Z_{i}=W * X_{i} \\
        f_{i}=\log \left(\frac{{VAR}\left(Z_{i}\right)}{{sum}\left({VAR}\left(z_{i}\right)\right)}\right)
        \end{array}\right.
\end{equation}

The work flow of the system is shown as Figure 2, and the BCI competition IV Dataset 2a with relatively few dimensions is utilized for classification.
\begin{figure}[H]
\centering
\includegraphics[width=4.0in,height=1.8in]{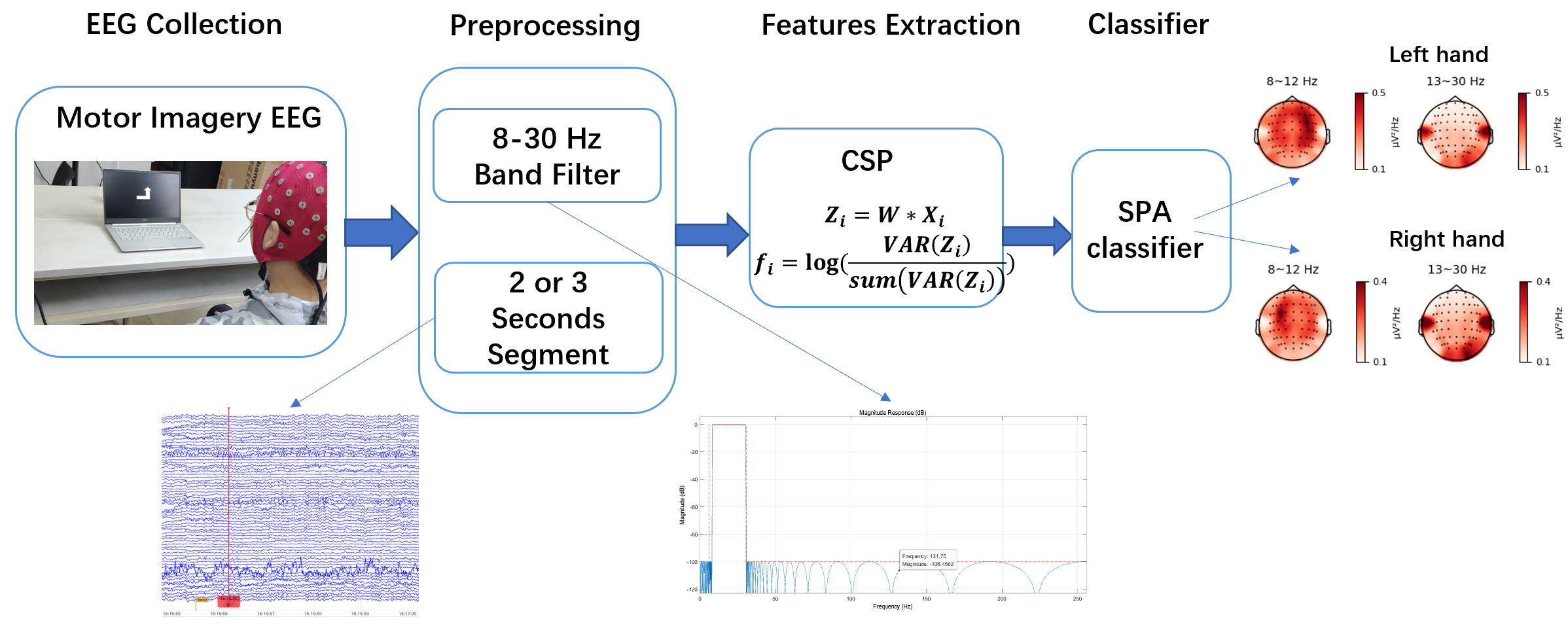}
\caption{Workflow of the system for EEG data classification}
\label{2} 
\end{figure}

\section{Principle and Implementation of SPA Classifier}

The basic idea of manifold learning is to assume that the dataset under study is uniformly sampled on a manifold structure in the high-dimensional data space, then identify the low-dimensional manifold structure hidden in the high-dimensional observation data space while keeping the neighborhood relationship between the data sets unchanged, and the nonlinear mapping relationship from the high-dimensional observation space to the low-dimensional embedding space is constructed for dimension reduction or visualization.

Manifold learning has the description \cite{de2003global}: given a dataset $X=\left\{x_{i}, i=1, \ldots, N\right\} \subset R^{m}$, assume that the samples in X are generated from a low-dimensional dataset Y through an unknown nonlinear transformation f, i.e.:
\begin{equation}
    x_{i}=f\left(y_{i}\right)+\varepsilon_{i}
\end{equation}
Among them, $\varepsilon_i$ denotes the noise, $y_{i} \in Y \subset R^{d}$, $d \gg m$, and $f: R^{d} \rightarrow R^{m}$ is the embedding map of $C^{\infty}$. Thus, the purpose of manifold learning is to get low-dimensional expressions based on the given observed dataset X:
\begin{equation}
    Y=\left\{y_{i}, i=1, \ldots, N\right\} \subset R^{d}
\end{equation}

Although many global and local methods have been proposed to identify low-dimensional embeddings, few of them sample the manifold information from the original data. Most global methods perform low-dimensional embedding by approximating the geodesic distance without the geodesic information of the unknown manifold, which can lead to bias \cite{freedman2002efficient}. The SPA method introduced in this paper mainly uses local manifold approximation and easy-to-fit spheres for local manifold approximation to classify the MI data. The following describes the details of the SPA classifier.

For the binary classification problem of MI, suppose there are two types of MI data with different types of motions, labeled as 1 and 2, respectively. The features of the MI data for the two different types of motions tend to approach two different manifolds $M_1$ and $M_2$, 
and both of which are embedded in $R^D$ with intrinsic dimension $p<D$. 
The two manifolds of the EEG signals might be highly nonlinear and complex, even have different curvatures or even gaps.

For a given test sample x, the distances between the sample and the two manifolds, denoted by $d_1$ and $d_2$, are first calculated, and then the sample is assigned to the group with a shorter distance. Actually, the two manifolds $M_1$ and $M_2$ are unknown, but the training data contains n different sample data and the corresponding labels. With this known information, it is possible to obtain an exact local approximation of the manifolds $M_1$ and $M_2$ in a high-order domain of a feature x to be processed. In this way, a wide variety of local approximations can be considered, and the limited training data can be used completely and efficiently.

\begin{table}[h]\footnotesize
\centering
\begin{tabular}{l}
    \toprule
    Algorithm 1: SPCA(Spherical Principal Components Analysis)algorithm  \\
    \midrule
    1 Input: $X=\left\{x_{i}\right\}_{i=1}^{n}$, Dimension of manifold p \\
    2 Output: Sphere $S_{P}(V, c, r)$ \\
    3 $V=\left(v_{1}, \ldots, v_{p+1}\right)$, Eigenvalues are in decreasing order, \\
    where $v_j$ is the j-th eigenvector of the covariance matrix \\
    4 $\xi_{i}=\bar{X}+V \bar{V}\left(X_{i}-\bar{X}\right)$ \\
    5 $c=-\frac{1}{2}\left\{\sum_{\mathrm{i}=1}^{n}\left(\xi_{i}-\bar{\xi}\right)\left(\xi_{i}-\bar{\xi}\right)^{\top}\right\}^{-1}\left\{\sum_{\mathrm{i}=1}^{n}\left(\xi_{i}^{\top} \xi_{i}-\frac{1}{n} \sum_{j=1}^{n} \xi_{j}^{\top} \xi_{j}\right)\left(\xi_{i}-\bar{\xi}\right)\right\}$ \\
    6 $r=\frac{1}{n} \sum_{\mathrm{i}=1}^{n}\left\|\xi_{i}-c\right\|$ \\
    \bottomrule  
\end{tabular}
\end{table}

The algorithm regards the local spherical approximation as the sphere which is a simple geometric object easy to fit, and the sphere provides a hyperplane generalization that can significantly improve accuracy by approximating local curvature. First, the center, radius and size of each sphere are optimized to provide the best local approximation. Then, the distances $d_1$ and $d_2$ between the test sample x and the two manifolds can be easily calculated based on the local spherical approximation. The sample to be classified by the distance to different manifolds embed in 3D space is shown as Figure 3.

\begin{figure}[H]
\centering
\includegraphics[width=4.00in,height=2.25in]{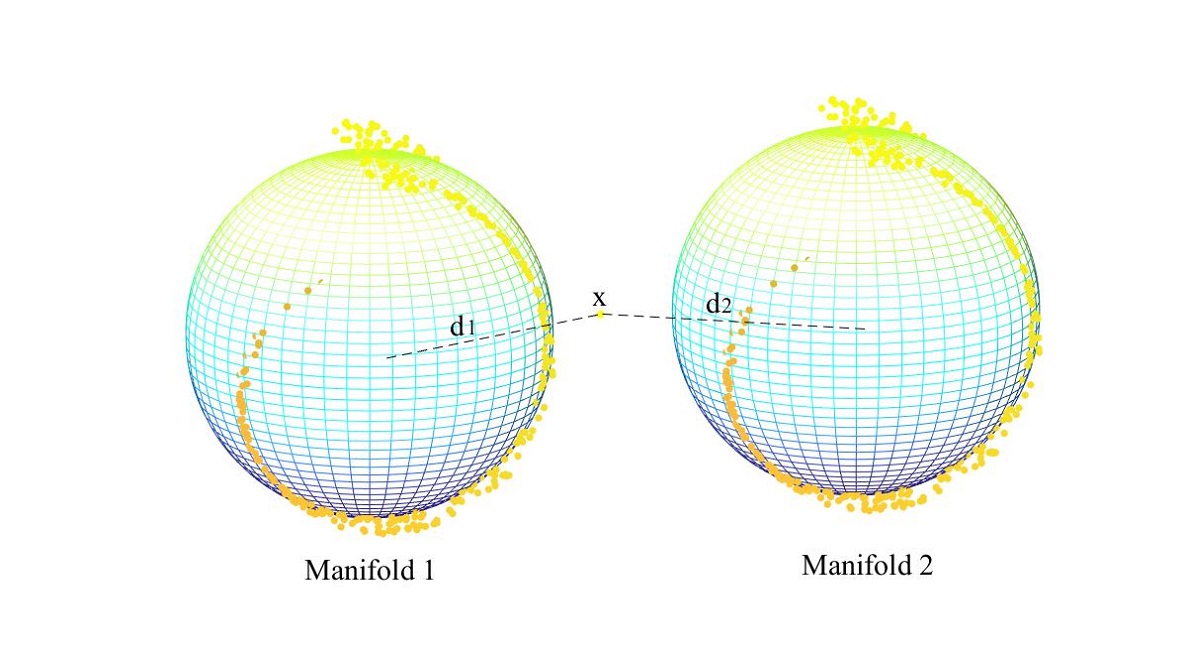}
\caption{Sample to be classified by the distance to different manifolds}
\label{3} 
\end{figure}

Let $X_{[k]}^{l}$ be the k-nearest neighbors between x samples with label $l$. The points are fitted to a sphere using the spherical principal component analysis (SPCA) algorithm to obtain a local manifold approximation $\widehat{M_{l}}$ around x in class l. Then, the value of $\mathrm{d}_{l}$ is approximated by $\widehat{\mathrm{d}_{l}}:=d\left(x, \widehat{M_{l}}\right)$, and the label y is selected as the value of $l$. The SPCA algorithm produces an estimated p-dimensional sphere $S_{P}(V, c, r)$ with center c and radius r in the subspace V when applied to data $X=\left\{x_{i}\right\}_{i=1}^{n}$.

\begin{table}[h]\footnotesize
	\centering
	\begin{tabular}{l}
    \toprule  
    Algorithm 2: SPA(SPherical Approximation)algorithm\\
    \midrule
    1 Input: training dataset$\left\{x_{i}, y_{i}\right\}_{i=1}^{n}$, parameters k and p,test data x\\
    2 Output: Prediction y of test data x\\
    3 Let L be the number of array\\
    4 for l=1:L\\
    5 Find the K-nearest neighbors of x under label l, denoted as $X_{[k]}^{l} \subset\left\{x_{i} \mid y_{i}=l\right\}$\\
    6 Compute the p-dimensional spherical approximation using Algorithm 1\\
    7 Compute the projection of x onto the sphere field S by :\\
    $({\mathord{\buildrel{\lower3pt\hbox{$\scriptscriptstyle\frown$}} 
    \over x} ^l}) = Pro{j^l}(x) = {c_l} + \frac{{{r_l}}}{{\left\| {{V_l}{V_l}^{\top}(x - {c_l})} \right\|}}{V_l}{V_l}^{\top}(x - {c_l})$\\
    8 Calculate the distance $\widehat{\mathrm{d}}_{l}$ between x and the sphere $S_{p}^{l}\left(V_{l}, c_{l}, r_{l}\right): \widehat{\mathrm{d}}_{l}=\left\|x-\widehat{x}^{\imath}\right\|$\\
    9 End\\
    10 Assign x to the group with the smallest distance: $y = \mathop {argmin}\limits_{l = 1, \ldots ,L} \widehat {{{\rm{d}}_l}}$\\
    \bottomrule  
\end{tabular}
\end{table}

The overall process of the SPA classifier is to first obtain the parameters V, c and r of the spherical classifier by Algorithm 1 (SPCA algorithm), and then implement the classification by Algorithm 2 (SPA algorithm).

The SPA classifier is simple to implement and efficient to perform, which could detect non-linear support differences between groups. The parameters to be tuned for the SPA classifier are the size of the local neighborhood k, and the dimension p that approximates the data denoising support. When these two parameters are fixed, the algorithm becomes fast. 

For the selection of parameters k and p, k and p are brought in from a certain range, and the highest accuracy is the accuracy of the test. For the selection of p, the interval is {1,2,3,4}; For the selection of k, the minimum value is set to 8, increasing in the order of 1, and the maximum value is set to the number of the small number of the two types of samples. If the number is too large, the maximum value is set to 46. If p is greater than the number of a small number of samples, the algorithm will fail.

The SPA classifier theoretically guarantees the impact of the growth of the number of training samples n on the classification performance. Theorem 1, which corresponds to the data in the presence of noise, is given below.

Theorem 1: Let $M_1$ and $M_2$ correspond to two Compact Riemannian Manifolds. Assume that $\left\{z_{i}\right\}_{i=1}^{n}{ }_{\sim}{ }^{i i d} \rho$, $supp(\rho)=M_{1} \cup M_{2}$, $z_{i} \in M_{y i}$ and $x_{i}=z_{i}+\epsilon_{i}$, where $\epsilon_{i} \sim N\left(0, \sigma^{2} I_{D}\right)$. Given a test sample x with label y, let the SPA classifier obtain the predicated label  $\widehat{y_{n}}$. Let $B_{\delta}(M):=\{x \mid \mathrm{d}(x-n)<\delta\}$ and $\delta>0$, then:

\begin{tiny} 
\begin{equation}
\lim _{n \rightarrow \infty} {P}\left(y \neq \widehat{y_{n}}\right) \leq \rho\left(B_{\delta}\left(M_{1}\right) \cap B_{\delta}\left(M_{2}\right)\right)+\exp \left\{-\frac{\delta^{2}}{8 \sigma^{2}}+\frac{D}{2} \log \left(\frac{\delta^{2}}{4 \sigma^{2}}\right)-\frac{D}{2}(\log (D)-1)\right\}
\end{equation}
\end{tiny} 

In theorem 1, the data in class $l$ are distributed around $M_{l}$ as Gaussian noise. This theorem shows that as the size of training sample n increases, the probability of the algorithm producing an error class label is gradually bounded by:

\begin{equation}
    \rho\left(B_{\delta}\left(M_{1}\right) \cap B_{\delta}\left(M_{2}\right)\right)+\exp \left\{-\frac{\delta^{2}}{8 \sigma^{2}}+\frac{D}{2} \log \left(\frac{\delta^{2}}{4 \sigma^{2}}\right)-\frac{D}{2}(\log (D)-1)\right\}
\end{equation}

When the noise level attenuates to 0, i.e., $\sigma \rightarrow 0$. Let $\delta=\sqrt{\sigma} \rightarrow 0$, so the first term converges to the $\rho\left(M_{1} \cap M_{2}\right)$, which is the intersection region between $M_1$ and $M_2$ are assigned by the probability p. Since $\frac{\delta^{2}}{\sigma^{2}} \rightarrow \infty$, the second term converges to 0, which is the case that the noise is zero. The probability of correct classification by the algorithm is ultimately greater than $1-\rho\left(M_{1} \cap M_{2}\right)$. As the deduction, the limit is 1 when $\rho\left(M_{1} \cap M_{2}\right)=0$, which means the SPA classifier has perfect classification performance when it has enough training samples, as long as the classes are geometrically separable. Theorem 1 takes into account the noise, which is more practical in most applications. $\frac{\delta^{2}}{\sigma^{2}}$  can be considered as the signal-noise ratio. The larger the $\frac{\delta^{2}}{\sigma^{2}}$ , the better the performance of the SPA classifier.

\section{Results Evaluation and Discussion}

The Accuracy \cite{xu2019deep} denotes the percentage of the number of correctly classified samples to the total number of samples, and is calculated as:

\begin{equation}
Accuracy=\frac{TP+TN}{TP+TN+FP+FN} \times 100 \%
\end{equation}

In the above formula, True Positive (TP) indicates the number of samples with positive category and positive classification prediction. True Negative (TN) indicates the number of samples with negative category and negative prediction. False Negative (FN) indicates the number of samples with positive category and negative prediction. And False Positive (FP) indicates the number of samples with negative category and positive prediction.

Algorithm evaluation: The algorithm proposed in this paper is evaluated by comparing the CSP+SPA algorithm with the following five algorithms.

\begin{enumerate}[(1)]
\item CSP+LDA: Motion imagery classification using CSP and LDA algorithm \cite{martinez2001pca}.
\item CSP+SVM: Motion imagery classification using CSP and SVM algorithm \cite{sun2010line}.
\item MDRM: The minimum distance to the Riemannian mean is used for classification on high-dimensional Riemannian manifolds \cite{barachant2011multiclass}.
\item TS+LDA: LDA was applied to the high-dimensional tangent space for classification \cite{barachant2011multiclass}.
\item TS+SVM: SVM was applied to the high-dimensional tangent space for classification \cite{liang2020calibrating}.
\end{enumerate}

Parameter Settings: In the TS+LDA algorithm, since the tangent space is $m=n \times(n+1) / 2$-dimensional space, there may be a problem that the sample dimension exceeds the number of trials per class, and the regularized classification algorithm is usually used to solve this problem. In the TS+SVM algorithm, the radial basis function kernel (RFB) is selected as the SVM kernel\cite{sun2010line}.

The following two experiments are designed for the BCI competition data, Experiment 1 for the normal number of samples and Experiment 2 for the small training dataset.

Since cross-validation allows testing the model in the training phase, in this paper, the performance of the proposed algorithm is first tested on the BCI IV Dataset 2a using a 10-fold cross-validation procedure. The BCI IV Dataset 2a is randomly divided into 10 subsets of equal size. In each run, 9 subsets are elected for training and 1 subset is left for testing.


\begin{table}[h]\footnotesize
    \caption{10-fold cross-validation results (\%)}			
	\centering
	\begin{tabular}{ccccccc}

    \toprule  
    \multirow{2}{*}{Subject}  & \multicolumn{6}{c}{Method}  \\ \cline{2-7}
                    & TS+LDA  & TS+SVM & MDRM  & CSP+SPA & CSP+SVM & CSP+LDA \\ \midrule  
	S01             & 84.29   & 86.07  & 80.71 & \textbf{88.93}   & 87.86   & 82.14\\  
    S02             & 59.64   & 60.36  & 54.29 & \textbf{70.36}   & 58.57   & 59.29\\  
    S03             & 96.43   & 96.07  & 91.07 & \textbf{97.50}   & 93.93   & 94.29\\  
    S04             & 81.07   & 82.50  & 76.07 & \textbf{83.93}   & 76.79   & 75.71\\  
    S05             & 59.64   & 60.00  & 57.50 & \textbf{71.43}   & 61.07   & 60.71\\  
    S06             & 69.29   & 68.93  & 68.21 & \textbf{73.93}   & 68.93   & 68.21\\  
    S07             & 80.00   & 81.07  & 75.36 & \textbf{90.36}   & 83.57   & 82.14\\  
    S08             & 95.00   & 94.64  & 92.14 & \textbf{98.57}   & 95.36   & 94.64\\  
    S09             & 86.07   & 85.36  & 77.14 & \textbf{86.43}   & 82.50   & 81.79\\  
    Mean            & 79.05   & 79.44  & 74.72 & \textbf{84.60}   & 78.73   & 77.66\\  
    Std             & ($\pm $)12.85   & ($\pm $)12.70  & ($\pm $)12.34 & \textbf{($\pm $)10.04}   & ($\pm $)12.68   & ($\pm $)12.20\\  
    \bottomrule  
\end{tabular}
\end{table}

BCI IV Dataset 2a is a four-category dataset, and this paper focuses on extracting the motion imagery of the left hand and right hand in the dataset. The classification accuracies of all algorithms are given in Table 1, and the classification accuracy of CSP+SPA is the highest on each subject. CSP+SPA, TS+LDA and TS+SVM not only have higher average classification accuracy but also have a lower standard deviation of classification accuracy. Among them, CSP+SPA has the lowest standard deviation, which indicates that the proposed SPA classifier is more robust to the variance of subjects than the other algorithms. The above results show that the SPA classifier works well for classifying the data of BCI IV Dataset 2a, which reflects the effectiveness of the algorithm.

For many reasons, the training set available in BCI applications is usually quite small. Reducing the number of training trials required for a specific task is an important goal for BCI, so the experiment is set up to evaluate the performance of the SPA classifier on small datasets. In the experiment, the training and testing samples of each subject are aggregated into a total sample set for that subject, from which different proportions of samples are selected for the experiment. 1/2, 1/3, 1/6, 1/12 of the BCI IV Dataset 2a (i.e., 144, 96, 48, and 24 trials randomly selected) were used as training samples in the evaluation, and 20 replicate experiments were conducted each time. Table 2 records the average classification accuracy with 1/2 and 1/6 of the total data as the training dataset. As the training sample size decreases from 1/2 to 1/6, the performance of all algorithms decreases. However, compared to other methods, the proposed method demonstrates less performance degradation and has the highest classification accuracy in most subjects.

\begin{table}[h]\tiny
    \caption{Accuracy in 1/2 and 1/6 training sets results (\%)}	
	\centering
    \setlength{\tabcolsep}{1.8mm}{
	\begin{tabular}{cccccccccccc}
    \toprule  
    \multirow{2}{*}{\makecell[c]{Sample\\Number}} &\multirow{2}{*}{Method} &\multirow{2}{*}{Mean} & \multicolumn{9}{c}{subject}  \\ \cline{4-12}
    &  &  &S01	&S02    &S03    &S04    &S05    &S06    &S07    &S08    &S09    \\ \midrule
    \multirow{6}{*}{\makecell[c]{1/2\\Training\\Sample\\(144\\trials)}} 
                         &TS+LDA	&75.65	&82.08	&56.84	&92.74	&\textbf{76.11}	&57.08	&64.72	&76.49	&92.15	&\textbf{82.60}\\
	                     &TS+SVM	&74.57	&80.20	&55.83	&92.66	&73.84	&56.36	&63.64	&74.88	&91.67	&82.02\\
	                     &MDRM	    &71.32	&81.25	&47.78	&90.38	&66.18	&52.22	&60.07	&75.94	&91.32	&76.74\\
	                     &CSP+SPA	&\textbf{77.94}	&\textbf{85.69}	&\textbf{60.59}	&\textbf{94.83}	&75.73	&\textbf{61.28}	&\textbf{65.49}	&\textbf{81.77}	&\textbf{95.76}	&80.35\\
                         &CSP+SVM	&75.25	&84.39	&54.86	&93.96	&72.36	&53.91	&63.47	&79.67	&93.91	&80.73\\
                         &CSP+LDA	&74.15	&78.92	&55.52	&92.99	&72.19	&55.03	&63.54	&78.33	&91.46	&79.38\\
                         \hline
    \multirow{6}{*}{\makecell[c]{1/6\\Training\\Sample\\(48\\trials)}}
                         &TS+LDA	&70.26	&77.21	&53.21	&89.40	&66.10	&53.00	&\textbf{60.13}	&66.02	&88.19	&79.10\\
	                     &TS+SVM	&67.90	&72.62	&55.27	&85.60	&60.51	&54.46	&58.31	&60.01	&85.33	&78.94\\
                         &MDRM	    &68.64	&79.15	&49.40	&89.90	&58.54	&49.67	&57.56	&67.77	&88.50	&77.31\\
                         &CSP+SPA	&\textbf{72.79}	&\textbf{80.25}	&\textbf{56.96}	&\textbf{91.50}	&\textbf{68.52}	&\textbf{56.42}	&59.23	&\textbf{74.27}	&\textbf{91.94}	&76.00\\
                         &CSP+SVM	&71.35	&79.16	&54.75	&90.77	&64.07	&51.61	&58.80	&71.35	&91.34	&\textbf{80.28}\\
                         &CSP+LDA	&65.62	&67.71	&51.67	&84.04	&61.52	&51.48	&55.48	&66.23	&83.79	&68.67\\
    \bottomrule  
\end{tabular}}
\end{table}

To further analyze the effect of the number of training samples on the performance of different subjects, the average classification accuracy of different subjects in the BCI IV Dataset 2a with the training set percentage of 1/2, 1/3, 1/6 and 1/12 is given. The line charts of the well-performing subject S08 and the poor-performing subject S06 are given below.

\begin{figure}[H]
\centering
\includegraphics[width=3.00in,height=3.00in]{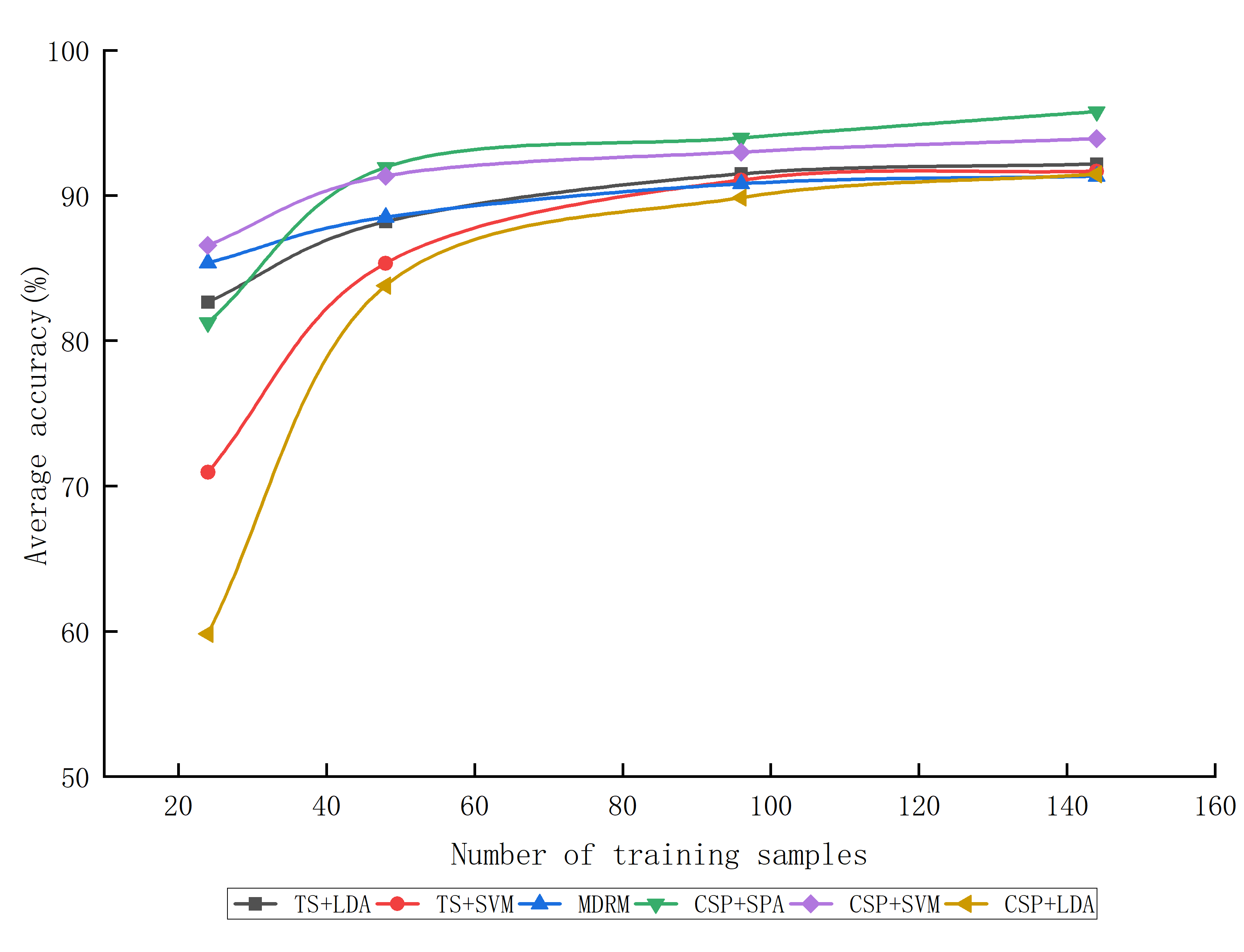}
\caption{Classification accuracy of S08}
\label{4} 
\end{figure}

\begin{figure}[H]
\centering
\includegraphics[width=3.00in,height=3.00in]{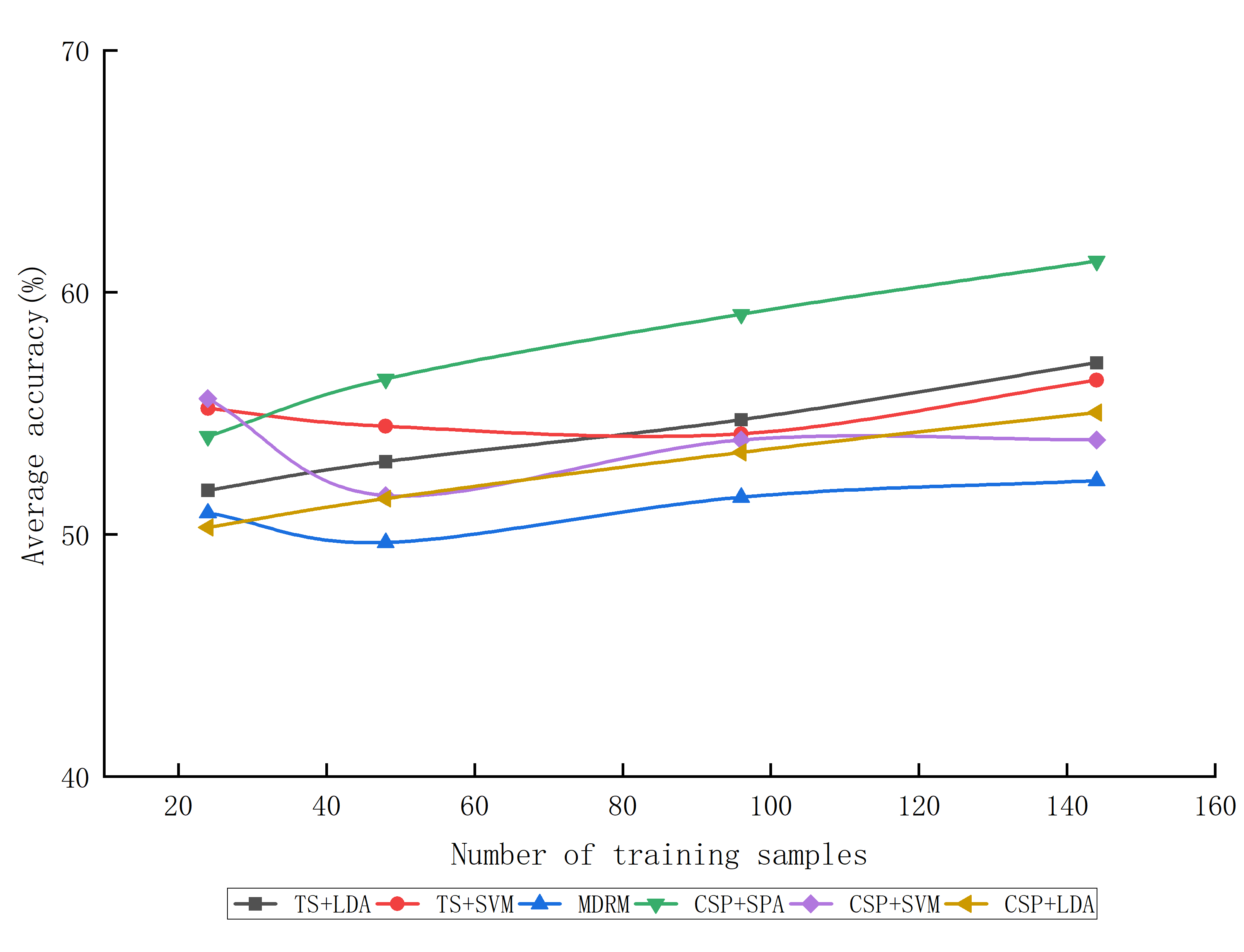}
\caption{Classification accuracy of S05}
\label{5} 
\end{figure}

As can be seen in Figure 2, the classification accuracy of the SPA classifier and the other algorithms on the well-performing subject S08 increases as the proportion of the training set increases. It is also obvious from the figure that the SPA classifier has the best performance in different proportions of the datasets. In Figure 3, there are cases where the classification accuracy increases despite a small proportion, which is due to the randomness of the poor performance of the subjects.

\section{Conclusion and Future Research}

MI does not rely on any external visual or auditory stimuli, which facilitates the implementation of portable BCI systems. It has broad application prospects in the fields of neural engineering, robotics, and rehabilitation engineering. Our work proposed a new classifier based on the spherical local manifold approximation, and optimized it for the best size of the local neighborhoods and the dimension of manifold facing the application of BCI systems.

To evaluate the performance of the proposed classification method, it runs on the 2008 BCI competition dataset. The experimental results demonstrate the effectiveness of the method in decoding and classifying two types of MI tasks, and the proposed method has higher decoding accuracy and shows strong robustness for small training datasets compared to other algorithms. The future work will focus on applying and improving the SPA classifier on the online rehabilitation devices for clinical experiments.


\bibliography{mybibfile}

\begin{thebibliography}{10}
\expandafter\ifx\csname url\endcsname\relax
  \def\url#1{\texttt{#1}}\fi
\expandafter\ifx\csname urlprefix\endcsname\relax\def\urlprefix{URL }\fi
\expandafter\ifx\csname href\endcsname\relax
  \def\href#1#2{#2} \def\path#1{#1}\fi

\bibitem{chaudhary2016brain}
U.~Chaudhary, N.~Birbaumer, A.~Ramos-Murguialday, Brain--computer interfaces
  for communication and rehabilitation, Nature Reviews Neurology 12~(9) (2016)
  513.

\bibitem{lazarou2018eeg}
I.~Lazarou, S.~Nikolopoulos, P.~C. Petrantonakis, I.~Kompatsiaris, M.~Tsolaki,
  Eeg-based brain--computer interfaces for communication and rehabilitation of
  people with motor impairment: a novel approach of the 21st century, Frontiers
  in human neuroscience 12 (2018) 14.

\bibitem{khan2014decoding}
M.~J. Khan, M.~J. Hong, K.-S. Hong, Decoding of four movement directions using
  hybrid nirs-eeg brain-computer interface, Frontiers in human neuroscience 8
  (2014) 244.

\bibitem{wang2006common}
Y.~Wang, S.~Gao, X.~Gao, Common spatial pattern method for channel selelction
  in motor imagery based brain-computer interface, in: 2005 IEEE engineering in
  medicine and biology 27th annual conference, IEEE, 2006, pp. 5392--5395.

\bibitem{brunner2007spatial}
C.~Brunner, M.~Naeem, R.~Leeb, B.~Graimann, G.~Pfurtscheller, Spatial filtering
  and selection of optimized components in four class motor imagery eeg data
  using independent components analysis, Pattern recognition letters 28~(8)
  (2007) 957--964.

\bibitem{lin2006frequency}
Z.~Lin, C.~Zhang, W.~Wu, X.~Gao, Frequency recognition based on canonical
  correlation analysis for ssvep-based bcis, IEEE transactions on biomedical
  engineering 53~(12) (2006) 2610--2614.

\bibitem{krusienski2008toward}
D.~J. Krusienski, E.~W. Sellers, D.~J. McFarland, T.~M. Vaughan, J.~R. Wolpaw,
  Toward enhanced p300 speller performance, Journal of neuroscience methods
  167~(1) (2008) 15--21.

\bibitem{naseer2015decoding}
N.~Naseer, K.-S. Hong, Decoding answers to four-choice questions using
  functional near infrared spectroscopy, Journal of Near Infrared Spectroscopy
  23~(1) (2015) 23--31.

\bibitem{foldes2017altered}
S.~T. Foldes, D.~J. Weber, J.~L. Collinger, Altered modulation of sensorimotor
  rhythms with chronic paralysis, Journal of neurophysiology 118~(4) (2017)
  2412--2420.

\bibitem{pichiorri2015brain}
F.~Pichiorri, G.~Morone, M.~Petti, J.~Toppi, I.~Pisotta, M.~Molinari,
  S.~Paolucci, M.~Inghilleri, L.~Astolfi, F.~Cincotti, et~al., Brain--computer
  interface boosts motor imagery practice during stroke recovery, Annals of
  neurology 77~(5) (2015) 851--865.

\bibitem{martens2010generative}
S.~Martens, J.~Leiva, A generative model approach for decoding in the visual
  event-related potential-based brain--computer interface speller, Journal of
  Neural Engineering 7~(2) (2010) 026003.

\bibitem{galan2008brain}
F.~Gal{\'a}n, M.~Nuttin, E.~Lew, P.~W. Ferrez, G.~Vanacker, J.~Philips,
  J.~d.~R. Mill{\'a}n, A brain-actuated wheelchair: asynchronous and
  non-invasive brain--computer interfaces for continuous control of robots,
  Clinical neurophysiology 119~(9) (2008) 2159--2169.

\bibitem{tangermann2008playing}
M.~Tangermann, M.~Krauledat, K.~Grzeska, M.~Sagebaum, B.~Blankertz,
  C.~Vidaurre, K.-R. M{\"u}ller, Playing pinball with non-invasive bci., in:
  NIPS, 2008, pp. 1641--1648.

\bibitem{fukuma2015closed}
R.~Fukuma, T.~Yanagisawa, S.~Yorifuji, R.~Kato, H.~Yokoi, M.~Hirata, Y.~Saitoh,
  H.~Kishima, Y.~Kamitani, T.~Yoshimine, Closed-loop control of a
  neuroprosthetic hand by magnetoencephalographic signals, PLoS One 10~(7)
  (2015) e0131547.

\bibitem{nijholt2009turning}
A.~Nijholt, D.~P.-O. Bos, B.~Reuderink, Turning shortcomings into challenges:
  Brain--computer interfaces for games, Entertainment computing 1~(2) (2009)
  85--94.

\bibitem{jackson2001adaptive}
Q.~Jackson, D.~A. Landgrebe, An adaptive classifier design for high-dimensional
  data analysis with a limited training data set, IEEE Transactions on
  Geoscience and Remote Sensing 39~(12) (2001) 2664--2679.

\bibitem{freedman2002efficient}
D.~Freedman, Efficient simplicial reconstructions of manifolds from their
  samples, IEEE transactions on pattern analysis and machine intelligence
  24~(10) (2002) 1349--1357.

\bibitem{xie2016motor}
X.~Xie, Z.~L. Yu, H.~Lu, Z.~Gu, Y.~Li, Motor imagery classification based on
  bilinear sub-manifold learning of symmetric positive-definite matrices, IEEE
  Transactions on Neural Systems and Rehabilitation Engineering 25~(6) (2016)
  504--516.

\bibitem{barachant2010riemannian}
A.~Barachant, S.~Bonnet, M.~Congedo, C.~Jutten, Riemannian geometry applied to
  bci classification, in: International conference on latent variable analysis
  and signal separation, Springer, 2010, pp. 629--636.

\bibitem{barachant2011multiclass}
A.~Barachant, S.~Bonnet, M.~Congedo, C.~Jutten, Multiclass brain--computer
  interface classification by riemannian geometry, IEEE Transactions on
  Biomedical Engineering 59~(4) (2011) 920--928.

\bibitem{tuzel2008pedestrian}
O.~Tuzel, F.~Porikli, P.~Meer, Pedestrian detection via classification on
  riemannian manifolds, IEEE transactions on pattern analysis and machine
  intelligence 30~(10) (2008) 1713--1727.

\bibitem{li2019classification}
D.~Li, D.~B. Dunson, Classification via local manifold approximation, arXiv
  preprint arXiv:1903.00985.

\bibitem{craik2019classification}
A.~Craik, A.~Kilicarslan, J.~L. Contreras-Vidal, Classification and transfer
  learning of eeg during a kinesthetic motor imagery task using deep
  convolutional neural networks, in: 2019 41st Annual International Conference
  of the IEEE Engineering in Medicine and Biology Society (EMBC), IEEE, 2019,
  pp. 3046--3049.

\bibitem{tangermann2012review}
M.~Tangermann, K.-R. M{\"u}ller, A.~Aertsen, N.~Birbaumer, C.~Braun,
  C.~Brunner, R.~Leeb, C.~Mehring, K.~J. Miller, G.~Mueller-Putz, et~al.,
  Review of the bci competition iv, Frontiers in neuroscience 6 (2012) 55.

\bibitem{asensio2013extracting}
J.~Asensio-Cubero, J.~Q. Gan, R.~Palaniappan, Extracting optimal tempo-spatial
  features using local discriminant bases and common spatial patterns for brain
  computer interfacing, Biomedical Signal Processing and Control 8~(6) (2013)
  772--778.

\bibitem{pfurtscheller2006mu}
G.~Pfurtscheller, C.~Brunner, A.~Schl{\"o}gl, F.~L. Da~Silva, Mu rhythm (de)
  synchronization and eeg single-trial classification of different motor
  imagery tasks, NeuroImage 31~(1) (2006) 153--159.

\bibitem{koles1990spatial}
Z.~J. Koles, M.~S. Lazar, S.~Z. Zhou, Spatial patterns underlying population
  differences in the background eeg, Brain topography 2~(4) (1990) 275--284.

\bibitem{ramoser2000optimal}
H.~Ramoser, J.~Muller-Gerking, G.~Pfurtscheller, Optimal spatial filtering of
  single trial eeg during imagined hand movement, IEEE transactions on
  rehabilitation engineering 8~(4) (2000) 441--446.

\bibitem{de2003global}
V.~De~Silva, J.~B. Tenenbaum, Global versus local methods in nonlinear
  dimensionality reduction, Advances in neural information processing systems
  (2003) 721--728.

\bibitem{xu2019deep}
G.~Xu, X.~Shen, S.~Chen, Y.~Zong, C.~Zhang, H.~Yue, M.~Liu, F.~Chen, W.~Che, A
  deep transfer convolutional neural network framework for eeg signal
  classification, IEEE Access 7 (2019) 112767--112776.

\bibitem{martinez2001pca}
A.~M. Martinez, A.~C. Kak, Pca versus lda, IEEE transactions on pattern
  analysis and machine intelligence 23~(2) (2001) 228--233.

\bibitem{sun2010line}
H.~Sun, Y.~Xiang, Y.~Sun, H.~Zhu, J.~Zeng, On-line eeg classification for
  brain-computer interface based on csp and svm, in: 2010 3rd International
  Congress on Image and Signal Processing, Vol.~9, IEEE, 2010, pp. 4105--4108.

\bibitem{liang2020calibrating}
Y.~Liang, Y.~Ma, Calibrating eeg features in motor imagery classification tasks
  with a small amount of current data using multisource fusion transfer
  learning, Biomedical Signal Processing and Control 62 (2020) 102101.

\end{thebibliography}

\end{document}